\documentclass[a4paper]{article}

\usepackage{icrc2013}

\title{MAGIC discovery and multiwavelength observations of the BL Lac 1ES 1727+502}

\shorttitle{1ES 1727+502: MAGIC and multiwavelength observations}

\authors{
K. Berger$^{1}$, G. De Caneva$^{2}$, E. Lindfors$^{3}$, S. Lombardi$^{4}$, A. Stamerra$^{5}$, F. Tavecchio$^{4}$ 
for the MAGIC Collaboration and S. Buson$^{6}$ for the \textit{Fermi}-LAT Collaboration.
}

\afiliations{
$^1$ Delaware University, 19711 Newark, DE, USA \\
$^2$ Deutsches Elektronen-Synchrotron (DESY), D-15738 Zeuthen, Germany \\
$^3$ Finnish Centre for Astronomy with ESO (FINCA), University of Turku, Finland \\
$^4$ INAF National Institute for Astrophysics, I-00136 Rome, Italy \\
$^5$ Universit\`a  di Siena, and INFN Pisa, I-53100 Siena, Italy \\
$^6$ Universit\`a di Padova and INFN, I-35131 Padova, Italy \\
}

\email{berger.karsten@gmail.com, gessica.de.caneva@desy.de}

\abstract{Blazars, active galactic nuclei whose jet axis is pointed towards the observer, constitute the most numerous class of extragalactic very high energy (VHE, E$>$100\, GeV) $\gamma$-ray emitters. The MAGIC experiment, a system of two Imaging Atmospheric Cherenkov Telescopes located in the Canary Island of La Palma (Northern hemisphere), with an energy threshold of 50\,GeV, is a well suited experiment for observations of such objects. 
 Here we present the discovery of the BL Lac 1ES 1727+502 ($z=0.055$) as VHE source. This object was identified as a promising TeV candidate based on archival data and the observation that lead to this detection was not triggered by any high state alert in other wavebands. The MAGIC observations are complemented by other observations are lower frequencies: optical data from the KVA telescope, UV, optical and X-ray archival data taken with the instruments on board the Swift satellite and high energy (HE, 300\,MeV$< E <$ 100\,GeV) data from the \textit{Fermi}-LAT instrument. We studied the spectral energy distribution of 1ES 1727+502 and interpreted it with a one-zone synchrotron self-Compton model with parameters that are typical for this class of sources.}

\keywords{BL Lac objects, 1ES 1727+502, gamma rays, multiwavelegth.}

\begin{document}
\maketitle

\section{Introduction}

The BL Lac 1ES 1727+502 is a blazar, an active galactic nuclei (AGN) with a relativistic jet pointing towards the observer.   Currently there are about 50 very high energy (VHE, E$>$100\, GeV) $\gamma$-ray emitting blazars\footnote{http://tevcat.uchicago.edu/}, constituting the most numerous class of extragalactic sources in this energy range. 

1ES 1727+502, located at a redshift $z = 0.055$ \cite{deVaucouleurs}, was observed with the Whipple 10-m $\gamma$-ray telescope \cite{Horan} and with the MAGIC-I telescope \cite{albert08}, without leading to a significant detection. It is also a source of high energy (HE, 300\,MeV$< E <$ 100\,GeV) $\gamma$ rays. The spectra observed by the \textit{Fermi}-LAT experiment is hard (with spectral index 2.0 in the \textit{Fermi}-LAT first source catalogue \cite{abdo10} and 1.8 in the second \textit{Fermi}-LAT catalogue \cite{nolan}). This, combined with the better sensitivity of the MAGIC experiment achieved after the installation of the second telescope, motivated the MAGIC observations which led to the first detection of this source in the VHE band \cite{Mariotti,aleksis13}.


\section{MAGIC discovery}

The VHE $\gamma$-ray observations were performed with the MAGIC experiment, a system of two 
Cherenov telescopes located in the Canary Island of La Palma \cite{cortina}. 

1ES 1727+502 was observed during 14 nights between May 6th and June 20th 2011. A quality selection based on the rates has been performed, removing runs taken during adverse atmospherical conditions or with technical problems. The effective time of the final data sample, corrected for the dead time of the triggers and the readout, is 12.6 h. Part of the data were taken under moderate moonlight and twilight conditions and were analyzed together with the dark night data \cite{Britzger}. The source was observed at zenith angles between 22$^{\circ}$ and 50$^{\circ}$, in false-source tracking mode \cite{Fomin} alternating every 20 minutes between two sky positions at $0.4^{\circ}$ offset from the source. 

\begin{figure}[t]
  \centering
  \includegraphics[width=0.4\textwidth]{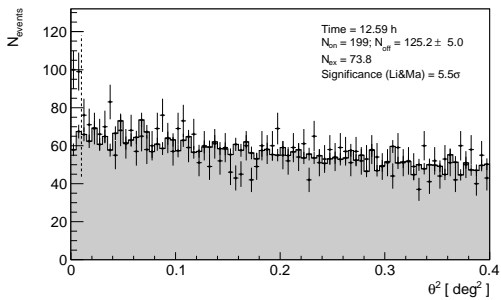}
  \caption{Distribution of the squared angular distance ($\theta^2$) between the source position and the reconstructed $\gamma$--ray direction for ON-source events (black points) and normalized OFF-source events (grey shaded area). The dashed line corresponds to the predefined region selected for the calculation of the significance of the detection. The respective statistics for ON and normalized OFF events are given in the figure.}
  \label{fig:theta2}
 \end{figure}

We have analysed the data with the standard MAGIC analysis framework \cite{Morajelo} with adaptations incorporating stereoscopic observations \cite{Lombardi}. As Fig.\ref{fig:theta2} shows, we have detected a signal with a significance of 5.5\,$\sigma$ (calculated with the formula 17 of \cite{LiMa}). The integral flux above 150\,GeV is $(2.1\pm0.4)\%$ of the Crab Nebula flux and the fitted position of the excess is consistent with the catalogue coordinates (RA: 17.47184$^\circ$, Dec: 50.21956$^\circ$ as in \cite{Ma}) within $ (0.032\pm0.015_{stat}\pm0.025_{sys}) ^\circ $, and thus compatible within the expected statistical and systematic errors \cite{aleksis12a}. Comparing the extension of the excess to the point spread function of MAGIC ($\sim 0.1^\circ$ \cite{aleksis12b}), the source appears to be point-like. The spectrum was unfolded, in order to take the finite energy resolution of the instrument into account. In the same procedure the flux was corrected for the absorption due to the interaction with low energy photons of the extragalactic background light using the model described in \cite{Dominguez}. The obtained differential flux can be described by a power law function dF$/$dE $ = f_0(\mathrm{E}/300\,\mathrm{GeV})^{-\Gamma}$ with the following values of the parameters: flux normalization  $f_0 = (9.6\pm2.5)\times 10^{-12}\,\mathrm{cm}^{-2}\,\mathrm{s}^{-1}\,\mathrm{TeV}^{-1}$ and spectral
index $\Gamma = (2.7\pm0.5)$. We estimate a 10\% additional systematic uncertainty in the measured flux compared to 
\cite{aleksis12a} due to the inclusion of moonlight and large zenith angle conditions in our data.

\begin{figure}[t]
  \centering
  \includegraphics[width=0.4\textwidth]{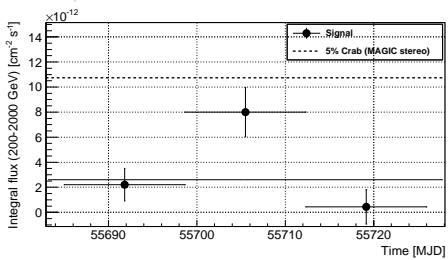}
  \caption{MAGIC light curve in the energy range from 200 GeV to 2 TeV. The Crab Nebula flux \cite{aleksis12a} scaled to 5\% is shown for comparison (dashed  line). The points correspond to the 14 days binned flux of 1ES 1727+502, and the error bars represent the statistical error only. The line represents the average flux during the entire observing period. The 
 probability of a constant flux is 0.6\% and the reduced $\chi^{2}$ with the number of degrees of freedom $n_{dof}$ of the fit assuming a constant flux is 10/2. }
  \label{fig:magicLC}
 \end{figure}

The VHE $\gamma$-ray light curve between 200\,GeV and 2\,TeV is shown in Fig.\ref{fig:magicLC}. In order to have a uniform distribution of days with observations in the bins and due to the weakness of the signal, a 14 day binning is applied starting from 2011 May 4. The emission is compatible with a constant flux of $(2.6\pm0.8)\times 10^{-12}\,\mathrm{cm}^{-2}\,\mathrm{s}^{-1}$. 
The relatively low probability of a constant flux (0.6\%, corresponding to a 2.5\,$\sigma$ rejection) might indicate variability below our detection threshold. The sparse binning and additional systematic errors due to moonlight and larger zenith angles can indeed fully explain this effect.

\section{Multiwavelength observations}

\subsection{\textit{Fermi}-LAT HE $\gamma$-ray observations}

\begin{figure}[t]
  \centering
  \includegraphics[width=0.4\textwidth]{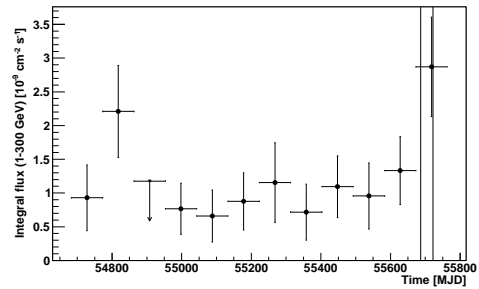}
  \caption{Light curve with a binning of three months of the \textit{Fermi}-LAT data between 1\,GeV and 300\,GeV. The downward pointing arrows correspond to a 95$\%$ upper limit. The vertical lines indicate the beginning and end of the MAGIC observing window in 2011. The emission is consistent with a constant flux, albeit a trend towards a higher flux in the last bin, partially coincident with the MAGIC observations, is evident.}
  \label{fig:fermiLC}
 \end{figure}

\begin{figure}[!h]
  \centering
  \includegraphics[width=0.4\textwidth]{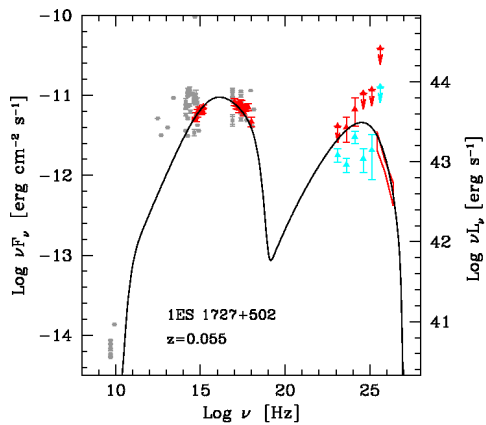}
  \caption{Multiwavelength spectral energy distribution fitted with a one-zone synchrotron self-Compton model \cite{Maraschi}). MAGIC observations, (red butterfly) have been corrected for the extragalactic background light absorption using the model \cite{Dominguez}. The data used for the fit (red triangles) are: optical from KVA, archival UV and optical from \textit{Swift}/UVOT, archival X-ray from \textit{Swift}/XRT, HE $\gamma$ rays from \textit{Fermi}-LAT (triangles, three months centred around the MAGIC observing period) and VHE $\gamma$ rays from MAGIC. We also show the 3-year LAT data (light blue triangles) and archival data (grey) from the ASI/ASDC archive ({\tt http://tools.asdc.asi.it/}).}
  \label{fig:mwsed}
 \end{figure}

1ES 1727+502 has been observed with the pair conversion Large Area Telescope (LAT) aboard \textit{Fermi} operating in the energy range from 20\,MeV up to  energies beyond 300\,GeV \cite{atwood,abdo10}. The data sample used for this analysis covers observations from August 5th, 2008 to August 5th, 2011. The analysis has been performed using the standard analysis tool {\it gtlike}, part of the \textit{Fermi} ScienceTools software package (version 09-27-01) available from the \textit{Fermi} Science Support Center (FSSC). The obtained light curve is presented in Fig.\ref{fig:fermiLC}. 
Possible variations in the source emission have been tested following the same likelihood method described in the second \textit{Fermi} catalogue \cite{nolan}. The results here obtained are consistent with a constant flux (test statistic $TS_{var}$=6 for 11 degrees of freedom), albeit a trend towards a higher flux in the last bin, partially coincident with the MAGIC observations, is evident. We also present in Fig.\ref{fig:mwsed} the spectrum obtained from three months of observations centred around the MAGIC observing period. Compared to the average flux, in the energy range from 300\,MeV to 300\,GeV, measured in three years of observations ($3.5\pm0.5\times 10^{-9}\;$ph cm$^{-2}$s$^{-1}$), the flux measured in the three months around the MAGIC observations is higher ($7.2\pm1.9\times 10^{-9}\;$ph cm$^{-2}$s$^{-1}$), while the spectral indices are similar (1.90$\pm$0.08 and 2.0$\pm$0.2 respectively). When performing the fit for the light curve and SED bins, the spectral indices of the sources were frozen to the best-fit values obtained from the time-independent analysis.

\subsection{\textit{Swift} XRT and UVTO observations}

 \begin{figure*}[!t]
  \centering 
  \includegraphics[width=0.3\textwidth, angle=270]{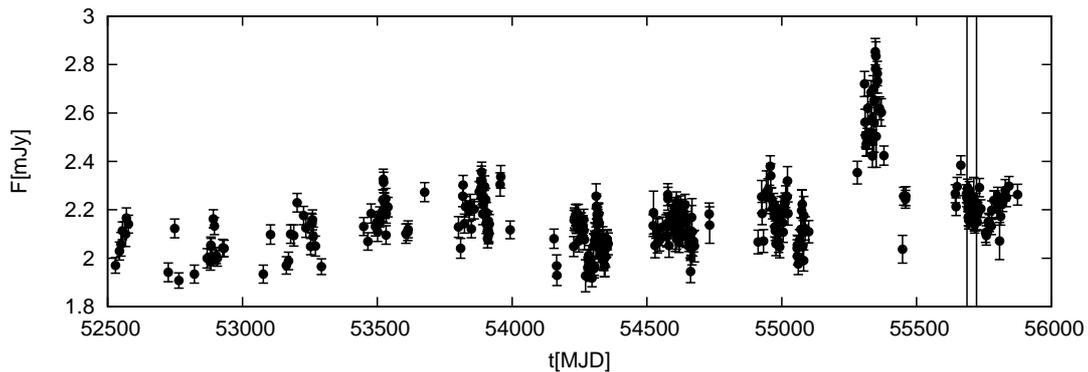}
  \caption{10 years light curve in the optical \textit{R}-band from the Tuorla blazar monitoring program. The contribution of the host galaxy ($1.25\pm0.06\,$mJy) has not been subtracted. Vertical lines indicate the beginning and end of the MAGIC observing window in 2011. See text for details.}
  \label{fig:kvaLC}
 \end{figure*}

The X-ray and UV/Optical data have been provided by two instruments aboard the \textit{Swift} Gamma-Ray Burst observatory:
the X-ray telescope (XRT, \cite{Burrows}) covering the 0.3$-$10\,keV energy band, and the UV/Optical Telescope (UVOT; \cite{Roming}) covering the 1800$-$6000\,$\AA{}$ wavelength range. Since there are no simultaneous \textit{Swift} observations during the MAGIC observing window, we have used archival data from April 5th and May 1st 2010. The data have been processed with standard procedures using the publicly available tools of the HEASoft package distributed by HEASARC. 

The data from \textit{Swift}/XRT have been fitted with a simple power law, in the range between 0.5$-$10\,keV. The flux is stable within $\sim 30\%$ during this period. \textit{Swift}/UVOT observations were performed during the same dates but only one of the observations, on April 5th, 2010 (MJD 55291.96182), contains all filters (\textit{V, B, U, W1, M2, W2}). We therefore used only this dataset for the compilation of the SED. It should be noticed that these archival \textit{Swift}/UVOT data were taken on 2010 April 5 when the optical flux was already increasing but before it reached the highest value, on 2010 May 31. Unfortunately there were no simultaneous observation with the KVA telescope but the \textit{R}-band SED point has a value of the flux, $4.93\pm0.2 \times 10^{-12}\,\mathrm{erg}\,\mathrm{cm}^{-2}\,\mathrm{s}^{-1}$, comparable to the spectral points obtained from \textit{Swift}/UVOT data. Consequently, the archival \textit{Swift}/UVOT can be regarded as representative of the baseline optical-UV flux and be included in the compilation of the multiwavelength SED. 

%

\subsection{KVA optical observations}

1ES 1727+502 has been observed continually in the optical \textit{R}-band as part of the Tuorla blazar monitoring program\footnote{http://users.utu.fi/kani/} for more than ten years, starting from 2002. The observations were carried out with the 1\,m Tuorla telescope and 35\,cm KVA telescope in La Palma. The brightness of the object was inferred from calibration stars in the same CCD-frames as 1ES 1727+502 using differential photometry and comparison star magnitudes from \cite{Fiorucci}. The magnitudes are converted to fluxes using the standard formula and values from \cite{Bessell}.

1ES 1727+502 has a bright host galaxy, contributing $>50\%$ to the flux in the optical \textit{R}-band \cite{Nilsson}. To derive the $\nu{\mathrm F}_\nu$ in the optical band, this contribution is subtracted from the measured flux and in addition the brightness was corrected for galactic absorption by \textit{R}=0.079\,mag \cite{Schlegel}. The average $\nu{\mathrm F}_\nu$ during the MAGIC observations corresponds to $(4.93\pm0.2) \times 10^{-12}\,\mathrm{erg}\,\mathrm{cm}^{-2}\,\mathrm{s}^{-1}$.

Overall, the source showed mainly quiescent behaviour (as shown in Fig.\ref{fig:kvaLC}) with the exception of an increased R-band flux starting in March-April 2010, with a peak value of $2.85\pm0.05$\,mJy on 2010 May 31, which exceeded the trigger criteria ($>$50\% above the long-term average) for MAGIC observations. However, the adverse atmospheric conditions forced us to discard the MAGIC data. The source had almost returned to its quiescent flux, $2.0 - 2.2$\,mJy, in September 2010 and remained in this state also during the MAGIC observations performed in 2011.

\subsection{Multiwavelength spectral energy distribution}

The quasi-simultaneous multiwavelength data described in the previous sections have been used for the compilation of the SED, which has been modelled with a one-zone SSC model \cite{Maraschi}. In this scenario, a blob of radius $R$ populated by relativistic electrons and filled with a tangled magnetic field of intensity $B$, is moving down the jet with a Doppler factor $\delta$. The electrons emit synchrotron radiation, producing the low-energy peak in the SED. The $\gamma$ rays are produced by the same electron population up-scattering the synchrotron photons, resulting in the second peak in the SED.

The electron spectrum is assumed to be described by $N(\gamma)=K \gamma^{-n_1}(1+\gamma/\gamma_b)^{n_1-n_2}$. The parameter values that give a good match between the SSC model and the SED data are: the Lorentz factors $\gamma_{min}=100, \gamma_{b}=3 \times 10^4, \gamma_{max}=6 \times 10^5$; the slopes $n_1=2, n_2=3.5$; and the electron density $K=8\times10^3\,$cm$^{-3}$.  The parameters that describe the astrophysical environment are the magnetic field $B=0.1\,$G, the radius $R=7\times 10^{15}\,$cm and the Doppler factor $\delta=15$ of the emitting region. 
These values are compatible with the values obtained with the sample analyzed in \cite{Tavecchio}.

\section{Conclusions}

The high frequency peaked BL Lac (HBL) 1ES 1727+502 shows little variability in the optical \textit{R}-band, is bright in the X-ray band, has a hard spectrum in the HE $\gamma$--ray band and, as shown in this paper, is visible in the VHE $\gamma$--ray range. The discovery of this source as VHE $\gamma$-ray emitter demonstrates the importance of combining data at different wavelengths, namely radio, optical, X-ray, and the recently opened {\it Fermi}--LAT energy range, to help identify potential VHE $\gamma$-ray emitters. The MAGIC detection indeed confirms the prediction made by \cite{Costamante,Donato} more than ten years ago, using X-ray, optical and radio data. Of the 33 sources in the list they compiled, 21 have been already detected. They predicted a flux of $0.7\times 10^{-12}\,\mathrm{cm}^{-2}\,\mathrm{s}^{-1}$ above 300$\,$GeV and we observed a flux a factor of two higher ($1.6\times 10^{-12}\,\mathrm{cm}^{-2}\,\mathrm{s}^{-1}$).

Furthermore it is also interesting to compare this result with the excess seen in the stacked AGN sample observed by MAGIC in mono mode\cite{aleksis11}. The spectral index measured for 1ES 1727+502 in the MAGIC energy range is compatible with the average spectral index of the stacked AGN sample: (2.7$\pm$0.5) compared to (3.2$\pm$0.5). Finally, when compared to the sample of all blazars detected in VHE $\gamma$-rays, its spectral index has the value of a typical BL Lac, while the flux is one of the lower fluxes detected so far 
\cite{Becerra12,Becerra13}.  

We have interpreted the emission with a single-zone SSC model and find that the model parameters are compatible with those obtained for other sources of the HBL class. We investigated the multiwavelength variability of the source. During MAGIC observations the source was in a quiescent state in the optical band, and the \textit{Fermi}-LAT data suggest (though not significantly) a flux enhancement during our observations compared to the three year averaged 
spectrum. 
We thus conclude that a study of the variability of this source, complemented with simultaneous multiwavelength observations, should be the focus of future 
observations. It will indeed help us in understanding not only the behaviour of this particular $\gamma$-ray emitter, but also the general features characterizing 
the HBL class of blazars.

\vspace*{0.5cm}
\footnotesize{{\bf Acknowledgments:}{We would like to thank the Instituto de Astrof\'{\i}sica de
Canarias for the excellent working conditions at the
Observatorio del Roque de los Muchachos in La Palma.
The support of the German BMBF and MPG, the Italian INFN, 
the Swiss National Fund SNF, and the Spanish MICINN is 
gratefully acknowledged. This work was also supported by the CPAN CSD2007-00042 and MultiDark
CSD2009-00064 projects of the Spanish Consolider-Ingenio 2010
programme, by grant DO02-353 of the Bulgarian NSF, by grant 127740 of 
the Academy of Finland,
by the DFG Cluster of Excellence ``Origin and Structure of the 
Universe'', by the DFG Collaborative Research Centers SFB823/C4 and SFB876/C3,
and by the Polish MNiSzW grant 745/N-HESS-MAGIC/2010/0.

The \textit{Fermi}-LAT Collaboration acknowledges support from a number of agencies and institutes for both development and the operation of the LAT as well as scientific data analysis. These include NASA and DOE in the United States, CEA/Irfu and IN2P3/CNRS in France, ASI and INFN in Italy, MEXT, KEK, and JAXA in Japan, and the K.~A.~Wallenberg Foundation, the Swedish Research Council and the National Space Board in Sweden. Additional support from INAF in Italy and CNES in France for science analysis during the operations phase is also gratefully acknowledged.The ICRC 2013 is funded by FAPERJ, CNPq, FAPESP, CAPES and IUPAP.}

\end{document}